\documentclass[aps,pra,twocolumn,superscriptaddress,amssymb,amsfonts,floatfix,showpacs]{revtex4}

\usepackage{graphicx,psfrag}
\usepackage{dcolumn}
\usepackage{bm}
\usepackage{mathbbol}
\usepackage{makeidx}               
\usepackage{graphicx}              
\usepackage[bookmarksnumbered,colorlinks,plainpages,citecolor=blue,urlcolor=blue]{hyperref}
\usepackage[rflt]{floatflt}
\usepackage{subfigure}
\usepackage{caption2}
\usepackage{latexsym}
\usepackage{amsmath}
\usepackage{amsfonts}
\usepackage{amssymb}
\usepackage{enumerate}

\newcommand{\uck}[1]{\o}

\newcommand{\ket}[1]{\mbox{$|#1\protect\rangle$}}

\newcommand{\bra}[1]{\mbox{$\protect\langle#1|$}}

\def\beq{\begin{equation}}
\def\eeq{\end{equation}}
\def\bea{\begin{eqnarray}}
\def\eea{\end{eqnarray}}

\begin{document}

\begin{titlepage}
\date{\today}

\title{Logical Entropy for Quantum States}
\author{Boaz Tamir}
\email{boaz.tamir@mail.huji.ac.il}
\affiliation{\mbox{Faculty of interdisciplinary studies, Bar Ilan University, Israel}}

\author{Eliahu Cohen}
\email{eliahuco@post.tau.ac.il}
\affiliation{\mbox{School of Physics and Astronomy, Tel Aviv University, Tel Aviv, Israel}}


\maketitle

\textbf{Abstract:}
The novel concept of quantum logical entropy is presented and analyzed. We prove several basic properties of this entropy with regard to density matrices. We hereby motivate a different approach for the assignment of quantum entropy to density matrices.

\vspace{1cm}

{\bf Key words:} Entropy; Logical entropy; Tsallis Entropy\\

\noindent
PACS numbers:03.67.-a; 89.70.Cf

\end{titlepage}


\section{Introduction and Motivation}
Entropy is an important measure of information in probability theory (Shannon entropy) and statistical mechanics (Gibbs entropy). A natural extension of these classical measures to the quantum realm is von Neumann entropy.
Despite being extremely useful and very common in the field of quantum information, the von Neumann entropy was  criticized on several different grounds \cite{Brukner1}\cite{Brukner2}\cite{Giraldi}; while classical entropy indicates one's ignorance about the system \cite{Cover}, quantum entropy is fundamentally different, reflecting the principle inaccessibility of information or the exitance of non-local correlations. Classical entropy is concerned with subjective/epistemic indefiniteness, while quantum entropy is concerned with objective/ontological indefiniteness \cite{Ellerman}. For this reason the non-additive Tsallis entropy \cite{Tsallis} and other measures such as \cite{Brukner1} were proposed.

Classical logical entropy was recently suggested in \cite{Ellerman} as a new information measure. It is a measure of the distinction between two partitions of a set. The set can be thought of as being originally fully distinct, while each partition collects together blocks whose distinctions are factored out. Each block represents the elements that are the same in some respect (they are formally associated with an equivalence relation on the set), hence the block is indefinite between the elements within it, but different blocks are still distinct from each other in that aspect.

We find this framework of partitions and distinction most suitable (at least conceptually) for describing the problems of quantum state discrimination, quantum cryptography and in general, for discussing quantum channel capacity. In these problems, we are basically interested in a distance measure between such sets of states, and this is exactly the kind of knowledge provided by logical entropy \cite{Ellerman}.
In this work we shall focus on the basic definitions and properties and leave other advanced topics for future research \cite{Next}.

Given a set $U$ and a partition $\pi = \{B\}$ of $U$ (where $B$ is the set of blocks in $U$), denote by dit$(\pi)$, the distinction (or `dit' for short) of the partition $\pi$, as the set of all pairs $(u,u')\in UxU$ such that $u$ and $u'$ are not in the same block $B$ of the partition $\pi$.
\indent Let the logical entropy $h(\pi)$ be defined as:
\begin{equation}
h(\pi) = \frac{|dit(\pi)|}{|UxU|}.
\end{equation}
\indent If $p_B = \frac{|B|}{|U|}$, then it is easy to see that
\begin{equation}
h(\pi) = 1- \sum_{B\in \pi} p_B^2.
\end{equation}
\noindent In other words, if we randomly draw two elements of $UxU$, then $h(\pi)$ is the probability that they are distinct, therefore it is a measure of the average distinction.
\indent Suppose $U=\{u_1,...,u_n\}$, and given a random variable with probabilities $\{p_1,...,p_n\}$, we can apply the above for the partition $1_U$ with $n$ one element-blocks $\{u_i\}$. Then
\begin{equation}
h(1_U) = 1- \sum_{i} p_i^2 = \sum p_i (1-p_i),
\end{equation}
\noindent therefore $h(1_U)$ is the probability to draw two different $u_i$'s consecutively.

Ellerman also defines logical relative entropy, logical conditional entropy and logical mutual information along the same lines \cite{Ellerman}.

In fact, logical entropy is rooted in the history of information theory. Polish Enigma crypto-analysts (and later Turing) used the term `repeat rate' \cite{Rejewski}. It is also a particular case of Tsallis entropy when $q=2$, and moreover resembles the information measure suggested by Brukner and Zeilinger \cite{Brukner1}\cite{Brukner2}.

Here we follow the standard methods in quantum information (e.g. \cite{Nielsen}) to extend the notion of logical entropy for describing quantum states. The set $U$ would be now a Hilbert space of a quantum system. This way we aim to generalize the results in \cite{QEllerman} and support them with formal proofs regarding quantum density matrices. The properties we prove hereby will hopefully shed new light on this intriguing informational measure.


Let us now extend the definition of logical entropy to the theory of quantum states:

\textbf{Definitions:}

Given a density matrix $\rho$, define the logical entropy $L(\rho)$ as:
\begin{equation}
L(\rho) = tr(\rho (1-\rho)).
\end{equation}
\noindent Let $\{ \lambda_i \}$ be the set of eigenvalues of $\rho$ then
\begin{equation}
L(\rho) = 1-\sum_i\lambda_i^2.
\end{equation}
In the following section we state and prove some of the basic properties of the quantum logical entropy.

\section{Properties of quantum logical entropy}

We will start with the definition of logical divergence and prove its non-negativity. The logical divergence takes the role of relative entropy in deriving some of the main results.\\

\textbf{Definition}

The logical divergence is defined as follows:
\begin{equation}
d(\rho||\sigma) = tr\rho(1-\sigma)- \frac{1}{2}tr(\rho(1-\rho))-\frac{1}{2}tr(\sigma(1-\sigma)).
\end{equation}

\textbf{Theorem II.1: Klein's inequality}
\begin{equation}
d(\rho||\sigma)\geq 0,
\end{equation}
\noindent with equality if and only if $\rho=\sigma$.

\textbf{Proof:} It is easy to verify that
\begin{equation}
d(\rho||\sigma) = \frac{1}{2} tr (\rho-\sigma)^2.
\end{equation}
\noindent

Observe now two very simple facts about Hermitian matrices:\

\textbf{Lemma II.1:}\\
Suppose $A$ is an Hermitian matrix, then:\\

\noindent a) $tr(A^2)\geq 0$\\
\noindent b) $tr(A^2)=0$ if and only if $A=0$\\

\indent\textbf{Proof of the lemma:} $A$ has real eigenvalues, hence the trace of $A^2$ is a sum of non-negative real numbers which equals zero if and only if all of them are zero.\\

This lemma completes the proof of the theorem. $\blacksquare$

\textbf{Theorem II.2: Basic properties:}

\noindent \textbf{1)} Logical entropy is non-negative and $L(\rho)=0$ for a pure state.

\noindent \textbf{2)} The maximal value of the logical entropy is $1-\frac{1}{d}$, where $d$ is the dimension of the Hilbert space. This value is the logical entropy of the maximally mixed state $I/d$.

\noindent\textbf{3)} Given a composite pure state $\rho^{A,B}$ on the space $(A,B)$ it follows that $L(\rho^A)=L(\rho^B)$.

\noindent\textbf{4)} If $\rho^{A,B} = \rho^A \otimes \rho^B$ then:

\begin{equation}
L(\rho^A\otimes \rho^B) = L(\rho^A) + L(\rho^B)- L(A)\cdot L(B)
\end{equation}


\textbf{Proof:}

\noindent \textbf{1)} For every density matrix, $tr \rho =1$ and $tr \rho^2 \le 1$ with equality if and only if $\rho$ is pure.


\noindent \textbf{2)} Use the above Klein inequality:
\begin{equation}
\frac{1}{2} tr(\rho(1-\rho))\leq tr \rho(1-I/d)- \frac{1}{2} tr I/d (1-I/d)
\end{equation}
\noindent Note now that $tr \rho(1-I/d) = tr I/d(1-I/d)= 1-\frac{1}{d}$

\noindent \textbf{3)} Immediate by the Schmidt decomposition, since $A$ and $B$ have the same orthonormal set of eigenvectors.

\noindent \textbf{4)} Write $\rho^A$ and $\rho^B$ in their diagonal form. Next use the following simple identity:
\begin{equation}
(1-x) + (1-y) - (1-x)(1-y) = 1-xy. \hspace{5mm}
\end{equation}
$~~~~~~~~~~~~~~~~~~~~~~~~~~~~~~~~~~~~~~~~~~~~~~~~~~~~~~~~~~~~~~~~~~~~~~~~~~\blacksquare$


In what follows we will sometime simplify the notation to denote $L(\rho^A)$ by $L(A)$.

\textbf{Definition:} We will say that $\rho^{A,B}$ is logical subadditive if:
\begin{equation}
L (A,B) \leq L(A)+ L(B)
\end{equation} \\
\textbf{Theorem II.3: Sufficient Conditions for logical subadditivity}

If $\rho^{A,B}$ can be diagonalized in a basis which is a tensor product of bases of $tr_A \rho^{A,B} =\rho^B$ and $tr_B \rho^{A,B} =\rho^A$ then $\rho^{A,B}$ is logical subadditive.

\textbf{Proof:} First write $\rho^{A,B}$ in its diagonal form and compute $L(A,B)$:
\begin{equation}
L(A,B)= 1- tr({\rho^{A,B}})^2 = 1- \sum_{i,j} a_{i,j,i,j}^2
\end{equation}
\noindent If we now trace out $A$ or $B$ we get diagonal matrices and we can easily compute $L(A)$ and $L(B)$:
\begin{equation}
L(A) =1- tr({\rho^A})^2 = 1- \sum_j(\sum_i a_{i,j,i,j})^2
\end{equation}
\begin{equation}
L(B)= 1- tr({\rho^B})^2 = 1- \sum_i(\sum_j a_{i,j,i,j})^2
\end{equation}

\noindent Note that the index $j$ ($i$) defines a partition $\pi_A$ ($\pi_B$) on the set
$\{a_{i,j,i,j}\}_{i,j}$ which is the diagonal of $\rho^{A,B}$. It is easy to see that the two partitions have no common block.

\noindent Hence we have to prove that:
\begin{equation}
\begin{array}{lcl}
1- \sum_j(\sum_i a_{i,j,i,j})^2 - \sum_i(\sum_j a_{i,j,i,j})^2 + \\
+ \sum_{i,j} a_{i,j,i,j}^2>0
\end{array}
\end{equation}

\noindent We can write the right hand side of the above equation as a sum of all products of pairs in $\pi_A$, or in $\pi_B$ (each pair appears exactly twice) and one copy of each of the $a_{i,j,i,j}^2$:
\begin{equation}
1- (\sum_{i,j} a_{i,j,i,j}^2 + \sum_j(\sum_{i,i'} a_{i,j,i,j}\cdot a_{i',j,i',j}+
\end{equation}
\[+\sum_i(\sum_{j,j'} a_{i,j,i,j}\cdot a_{i,j',i,j'}))\]

Clearly this expression is bigger than  $1- (\sum_{i,j} a_{i,j,i,j})^2$ which contains products of all pairs in the diagonal of $\rho^{A,B}$. However, $1- (\sum_{i,j} a_{i,j,i,j})^2=0$.  $\blacksquare$

\textbf{Theorem II.4: Logical entropy of a measured density matrix}

Let us assume a particular case of POVM defined by the projectors $P_i$, where $\sum_i P_i =1$ and $P_i P_j =\delta_{ij}P_j$. Let $\rho$ be a density matrix, and $\rho' = \sum_i P_i \rho P_i$ be the density matrix following the measurement, then:
\begin{equation}
L(\rho) \leq L(\rho').
\end{equation}
\textbf{Proof:}
By the Klein inequality $d(\rho||\rho')\geq 0$ we have:
\begin{equation}
\frac{1}{2} tr\rho(1-\rho) \leq tr\rho(1-\rho')- \frac{1}{2} tr\rho'(1-\rho').
\end{equation}

\noindent Therefore it is enough to show that $tr\rho(1-\rho')=tr\rho'(1-\rho')$.
Since $\sum_i P_i =1$ and $P_i P_j =\delta_{ij} P_j$ we have:
\begin{equation}
\begin{array}{lcl}
tr\rho(1-\rho')= tr \{(\sum P_i)\rho(1-\rho')(\sum P_i)\}= \\
=tr \sum P_i \rho(1-\rho')P_i,
\end{array}
\end{equation}
\noindent where in the last equality we have used the definition of the trace. However, $\rho'P_i = P_i\rho'$ and therefore:
\begin{equation}
tr\sum P_i \rho(1-\rho')P_i = tr \sum P_i \rho P_i(1-\rho') = tr\rho'(1-\rho').
\end{equation}
This completes the proof of the theorem. $\blacksquare$ \



The next theorem discusses the weighted sum of entropies on a given space.

\textbf{Theorem II.5: Concavity of logical entropy}

Let $\rho = \sum p_i \rho_i$, for $\{p_i\}_i$ some distribution and set $\overline{L(\rho)} = \sum p_i L(\rho_i)$ , then:

\noindent \textbf{a)} If $\rho_i$ have orthogonal support then:
\begin{equation}
\overline{L(\rho)}< L(\rho).
\end{equation}
\noindent \textbf{b)} In general:
\begin{equation}
\overline{L(\rho)}- L(p_i) < L(\rho) < \overline{L(\rho)}+ L(p_i).
\end{equation}
\noindent In other words $L(\rho)$ is in the $L(p_i)$ neighborhood of $\overline{L(\rho)}$, where $L(p_i)$ is the classical logical entropy of the distribution $\{p_i\}_i$.\\

\textbf{Proof of a):}
We will demonstrate the argument on two density matrices $\rho_1$ and $\rho_2$ having an orthogonal support. Set $\rho_1 = \sum_i p_i \ket{i}\bra{i} $, $\rho_2 = \sum_j q_j \ket{j}\bra{j} $
where $\ket{i}$ and $\ket{j}$ are two bases with orthogonal support, also set $\rho = \lambda \rho_1 + (1-\lambda) \rho_2$, where $0<\lambda<1$. Now:
\begin{equation}
\begin{array} {lcl}
\overline{L(\rho)} = \lambda L(\rho_1) + (1-\lambda) L(\rho_2) = \\
\lambda  ( 1- \sum_i p_i^2) + (1-\lambda) (1- \sum_j q_j^2)= \\
1-\lambda \sum_i p_i^2 - (1-\lambda)\sum_j q_j^2 .
\end{array}
\end{equation}
\noindent However,
\begin{equation}
L(\rho) = 1-\lambda^2 \sum_i p_i^2 - (1-\lambda)^2\sum_j q_j^2
\end{equation}
\noindent Therefore
\begin{equation}
\overline{L(\rho)} < L(\rho)
\end{equation}
\textbf{Proof of b):}
Consider $\rho^{A,B} = \sum_i p_i \rho_i\otimes \ket{i} \bra{i}$, so
$\rho^{A,B}$ is a sum of densities with an orthogonal support. From a) above and the logical subadditivity of $\rho^{A,B}$ (see theorem II.3 above) it follows that:
\begin{equation}
\overline{L(\rho)} \leq L (\rho^{A,B}) \leq L(A) + L(B)
\end{equation}
\noindent However $L(A)= L(\rho)$ and $L(B)= L(p_i)$, therefore:
\begin{equation}
\overline{L(\rho)} \leq L(\rho) +L(p_i),
\end{equation}
\noindent or
\begin{equation}
L(\rho) \geq \overline{L(\rho)} - L(p_i).
\end{equation}
This concludes the first part of b). It is left to show that
\begin{equation}
L(\rho)\leq \overline{L(\rho)} +L(p_i).
\end{equation}
\noindent We will first prove the above inequality for the case where the density is a sum of pure states:
\begin{equation}
\rho = \sum_i p_i \ket{\psi_i} \bra{\psi_i},
\end{equation}
\noindent where $\ket{\psi_i}\bra{\psi_i}$ are pure states of the system $A$. Consider the auxiliary pure state on $(A,B)$:
\begin{equation}
\begin{array} {lcl}
\ket{\eta} = \sum_i \sqrt{p_i} \ket{\psi_i} \otimes \ket{i} \\
\tilde{\rho} = \ket{\eta}\bra{\eta},
\end{array}
\end{equation}
\noindent where $\ket{i}$ is orthogonal in some system $B$. Then by theorem II.2.3 above:
\begin{equation}
L(\tilde{\rho}^B) = L(\tilde{\rho}^A) = L(\rho)
\end{equation}
\noindent By tracing out the system $A$ (note that $\ket{\psi_i}$ are not necessarily orthogonal) we get
\begin{equation}
\tilde{\rho}^{B} =\sum_{i,j} \sqrt{p_i p_j} \bra{\psi_j} \psi_i\rangle \ket{i}\bra{j}
\end{equation}
\noindent Measuring $B$ with the operators $P_i = \ket{i}\bra{i}$ we get:
\begin{equation}
\tilde{\rho}^{B'} = \sum_i p_i \ket{i} \bra{i}.
\end{equation}
\noindent By theorem II.4 above:
\begin{equation}
L(\tilde{\rho}^{B'}) = L(p_i) \geq L(\tilde{\rho}^B) = L(\rho).
\end{equation}
\noindent Therefore for $\rho$ which is a sum of pure states we have:
\begin{equation}
L(\rho)\leq L(p_i).
\end{equation}
Consider now the general case where:
\begin{equation}
\rho = \sum_i p_i \rho_i,
\end{equation}
\noindent and
\begin{equation}
\rho_i = \sum_j p_i^j \ket{e_i^j} \bra{e_i^j},
\end{equation}
\noindent where the vectors in $\{\ket{e_i^j}\}_j$ are orthogonal for each $i$. Hence
\begin{equation}
\rho = \sum_{i,j} p_i p_i^j \ket{e_i^j} \bra{e_i^j}.
\end{equation}
\noindent Here $\ket{e_i^j} \bra{e_i^j}$ are pure for all $i$ and $j$, and we can use the above to conclude:
\begin{equation}
L(\rho) \leq L(p_i p_i^j) = \sum_{i,j} p_i p_i^j (1-p_i p_i^j).
\end{equation}
\noindent We use now the simple fact that for $0\leq x_i, x_j\leq 1$:
\begin{equation}
1-x_i x_j \leq (1-x_i) + (1-x_j),
\end{equation}
\noindent therefore:
\begin{equation}
\begin{array} {lcl}
\sum_{i,j} p_i p_i^j (1-p_i p_i^j)\leq \sum_{i,j} p_i p_i^j (1-p_i)+ \\
+\sum_{i,j} p_i p_i^j (1-p_i^j)= L(p_i) + \sum_i p_i L(\rho_i),
\end{array}
\end{equation}
\noindent where we have used the orthogonality of the set of vectors $\{\ket{e_i^j}\}_j$ for each $i$. $\hspace{20mm} \blacksquare$

\textbf{Theorem II.6: The joint convexity of logical divergence}

The logical divergence $d(\rho||\sigma)$ is jointly convex.

\textbf{Proof:} We shall use the fact the $tr(\rho^2)$ is convex from the convexity of $x^2$ and the linearity of the trace to write:

\[ d(\lambda \rho_1 + (1-\lambda) \rho_2|| \lambda \sigma_1 + (1-\lambda) \sigma_2)=\] \[=tr((\lambda \rho_1 +(1-\lambda)\rho_2) -(\lambda\sigma_1 + (1-\lambda) \sigma_2))^2 =\]
\[= tr(\lambda(\rho_1-\sigma_1) + (1-\lambda)(\rho_2-\sigma_2))^2 \leq \]
\[\leq \lambda tr(\rho_1-\sigma_1)^2 + (1-\lambda) tr(\rho_2-\sigma_2)^2 =\]
\[= \lambda d(\rho_1||\sigma_1) +(1-\lambda) d(\rho_2||\sigma_2) \]

\noindent where the inequality is due to the convexity of $tr(x^2)$. This constitute the joint convexity.  $\blacksquare$



The following theorem states the fact that the divergence behaves as a metric. Tracing out a subspace only reduces the distance.\\

\textbf{Theorem II.7: The monotonicity of logical divergence}

Let $\rho^{A,B}$ and $\sigma^{A,B}$ be two density matrices, then:

\begin{equation}
d(\rho^A\otimes I/b||\sigma^A\otimes I/b) \leq d(\rho^{A,B}||\sigma^{A,B}),
\end{equation}
\noindent where $b$ is the dimension of $B$.

\textbf{Proof:} Observe that there is a set of unitary matrices $U_j$ over $B$ and a probability distribution $p_j$ such that:
\begin{equation}
\rho^A \otimes I/b = \sum_j p_j U_j \rho^{A,B} U_j^\dag
\end{equation}
\begin{equation}
\sigma^A \otimes I/b = \sum_j p_j U_j \sigma^{A,B} U_j^\dag,
\end{equation}
\noindent (see \cite{Nielsen} chapter 11). Now since $d(\rho||\sigma)$ is jointly convex on both densities, we can write:

\begin{equation}
d(\rho^A\otimes I/b||\sigma^A\otimes I/b) \leq \sum_{j} p_j  \cdot d(U_j \rho^{A,B} U_j^\dag||U_j \sigma^{A,B} U_j^\dag).
\end{equation}

Observe now that the divergence is invariant under unitary conjugation, and therefore the above sum is $d(\rho^{A,B}||\sigma^{A,B})$. $\hspace{5mm} \blacksquare$ \\

\noindent \section{Discussion}
Logical entropy might be more intuitive and useful than von Neumann entropy when analyzing specific quantum problems \cite{Zurek}\cite{Buscemi}. By its construction, logical entropy seems to reflect more naturally the objective indefiniteness of quantum mechanics. Two elements in the same set (according to some partition) are intrinsically indistinguishable.

We have shown some basic properties of the logical entropy for quantum states. These are very similar to the properties held by the standard von Neuman entropy. Note the small differences concerning the concavity of the entropy as stated in theorem II.5. We note that the logical entropy does not fulfill the strong subadditivity property which is an important feature of the standard von Neuman entropy. However, it was shown to possess a weaker version of subadditivity known as `firm subadditivity' \cite{Coles}. We suspect that the lack of this property might have fundamental role, e.g. while discussing the breakdown of sub-additivity in black holes \cite{AMPS}.

Being a measure of distinctions it is only natural to investigate channel capacity in terms of quantum dits. We expect that the language of quantum dits can simplify the proofs of channel properties. It would also be interesting to examine the use of logical entropy in the context of entanglement quantification in discrete/continuous systems.

\section{Acknowledgements}

We wish to thank Y. Neuman from Ben Gurion University of the Negev for numerous discussions on the topic of information as distinctions. We are also thankful for Judy Kupferman for helpful comments and discussions. E.C. was partially supported by Israel Science Foundation Grant No. 1311/14.

\section{References}

\end{document}